\documentclass[aps,prl,twocolumn,superscriptaddress,amsfonts,showpacs,floatfix]{revtex4}

\usepackage{amsmath,amssymb,bm,graphicx,epsfig,psfrag}

\usepackage[normalem]{ulem} 
\usepackage{multirow}

\newcommand{\Ru}{\mathrm{Re}}
\newcommand{\Rm}{\mathrm{Rm}}

\begin{document}

\title{Role of Cross Helicity in Cascade Processes of MHD turbulence}

\author{Irina Mizeva}
\author{Rodion Stepanov}
\author{Peter Frick}
\affiliation{Institute of Continuous Media Mechanics,
Academy of Sciences, 1 Korolyov St., Perm, 614013, Russia}

\date{Published in Doklady Physics, 2009, Vol. 54, No. 2, pp. 93–97.}
\maketitle

The cross helicity $H = <\vec v \cdot \vec b>$ characterizes the
level of correlation between pulsations of the magnetic field
$\vec b$ and the velocity field $\vec v$. In the ideal
three-dimensional magnetic hydrodynamics, it is an integral of
motion along with the total energy $E^T = E^v + E^b$, $E^v =
<|\vec v|^2/2>$, $E^b=<|\vec b|^2/2>$. The third integral of
motion is the magnetic helicity; but within the framework of this
study, we consider the fields in which the average magnetic
helicity is close to zero.

As developed turbulence is random process one can expect that, if there are no special
reasons, the developed turbulence of conducting fluid (the MHD
turbulence) should be characterized by a low level of cross-helicity.
Exactly such a situation is usually considered. Interest in cross
helicity arose after highly correlated pulsations of velocity and
magnetic field were found in the solar wind \cite{solarwind}.
Analysis of the energy and helicity evolution in a freely
decaying MHD turbulence showed that the helicity decays more
slowly than the energy; hence, the degree of correlation of fields
$\vec v$ and $\vec b$ determined by the correlation coefficient
$C=H/E^T$, can increase in time for the free decay
\cite{dobrowolny80}.

By itself, MHD turbulence gives the possibility of developing
various scenarios. The specificity of the conducting fluid
hydrodynamics is the possibility of occurrence of Alfven waves; it
is assumed they play a key role in the turbulent cascade, which
leads to the Iroshnikov-Kraichnan spectral law $E^b(k)\sim
E^v(k)\sim k^{-3/2}$ \cite{ir,kr}. In the simulations on the grid
$512^3$ \cite{muller_biskamp2000}, it was shown that no Kraichnan-Iroshnikov spectrum arises in the
noncorrelated turbulence without an external field, and the
turbulence with the spectrum close to $E^b(k)\sim E^v(k)\sim
k^{-5/3}$ is realized in the inertial interval. In
\cite{muller_grappin2005} the grid was expanded to $1024^3$ and
the turbulence was considered both with and without the
external field. For the MHD turbulence without the external field,
the "$-5/3$" law was confirmed and a significant anisotropy for
which the transverse pulsations follow the Iroshnikov–Kraichnan
law was revealed in the external field.

The problem of the cross-helicity effect on the forced MHD
turbulence was considered in \cite{gpl83} only in the context of
the Alfven scenario (i.e., the turbulence that gives the "$–3/2$"
spectrum without the cross-helicity source). On the basis of the
EDQNM approximation, it was shown that the system tends to the
steady state in which the correlation coefficient proves to be
much higher than the ratio of energy to cross-helicity input rates.
In this case, the energy spectrum becomes steeper.

The purpose of this work is to investigate the spectral properties
of the developed isotropic (non-Alfven) MHD turbulence stationary
excited by an external force, which also injects the cross
helicity into the flow simultaneously with the energy.

We consider MHD turbulence with the magnetic Prandtl number of order of unity. On the integral scale $L$, an external force acts with a
energy input rate equal to $\varepsilon$. The same forces inject a
certain cross-helicity input rate $\chi$  and
 no magnetic helicity. We assume  the
equidistribution of kinetic and magnetic energy $\delta
v_l^2\approx \delta b_l^2 \approx E_l$ within the limits of the
inertial range. Also in the inertial range the
energy flux is constant on any scale $l$ and equal to the
energy dissipation rate (the energy input rate)
 \begin{equation}
\frac{E_l}{t_l}=\varepsilon, \label{eq:eps}
\end{equation}
where $t_l$ is the characteristic exchange time. In the theory of
 isotropic turbulence this time is usually estimated as  the eddy turnover time
$\tau_l \approx {l}/{\delta v_l}$. In the case of
noncorrelated velocity and magnetic-field pulsations,
$H=0$ and this estimate can be accepted also for the
MHD turbulence in which nonlinear interactions dominate
instead of the Alfven waves.

The basic idea of further arguments is that the injected cross helicity
delays the spectral exchange (increases the time)
\begin{equation}
t_l=\frac{l}{\delta v_l}\xi_l = \tau_l \xi_l,  \label{eq:tl}
\end{equation}
and the delay coefficient $\xi_l$ is related to the correlation
level for the velocity and magnetic-field pulsations on
this scale. Hypothesis (\ref{eq:tl}) leads to the estimation of
energy pulsations on the scale $l$ in the form
\begin{equation}
\delta v_l^2 \approx (\varepsilon l \xi_l)^{2/3} .
\label{eq:en0}\end{equation}
In this case, the delay coefficient actually determines
the deviation from the Kolmogorov "4/5" law
\begin{equation}
 \xi_l  \approx  \frac{\delta v_l^3}{\varepsilon l},
\label{eq:qq}
\end{equation}
and Eq.~(\ref{eq:en0}) coincides with the Kolmogorov-Obukhov
law $\delta v_l^2 \approx (\varepsilon l)^{2/3}$ for $\xi_l=1$.

The delay of the cascade processes should lead to accumulation of the energy
of turbulent oscilations (in comparison
with the noncorrelated turbulence for the same
power source). The application of estimate (\ref{eq:en0}) to the
energy-transfer scale $l=L$ for which $v_L^2 \approx E$ gives
\begin{equation}
E \approx (\varepsilon L \xi_L)^{2/3}. \label{eq:eps2}
\end{equation}

The simplest assumption about the form of $\xi_L$ consists
in the fact that, on the scales of action of external
forces, the delay of the cascade transfer is determined
by the quantity $(1-\chi/\epsilon)$, which is a characteristic of noncorrelation
of perturbations introduced by external
forces. Taking into account the quadraticity of the terms
describing the processes of spectral transfer, we can
assume that $\xi_L \approx(1-\chi/\varepsilon)^{-2}$
which gives the estimate for
the average energy of the stationary forced turbulence
\begin{equation}
E\approx \frac{(\varepsilon L)^{2/3}}{(1-{\chi/{\varepsilon}})^{4/3}}.
\label{eq:En(hi)}
\end{equation}

Thus, the cross helicity blocks cascade energy transfer
and leads to energy accumulation in the system.
This accumulation proceeds until the vortex intensification
compensates the decreasing efficiency of nonlinear
interactions.

When assuming that the external forces inject
the cross helicity in the turbulence with the given flux
$\chi$, it is necessary also to accept the hypothesis that the
cross-helicity flux is constant over the spectrum for a
steady state, which gives
\begin{equation}
\frac{H_{l}}{t_l^{(\chi)}}=\chi, \label{eq:chi}
\end{equation}
at that, the cross-helicity-exchange time
$t_l^{(\chi)}=\tau_l \xi_{l}^{(\chi)}$, could not coincide with the energy-exchange time $t_l$.
Let $C_l$ be the correlation coefficient for the velocity
and magnetic-field pulsations on the scale $l$
\begin{equation}
C_l=\frac{<\delta v_l \delta b_l>}{\sqrt{<\delta v_l^2> <\delta
b_l^2>}}\approx \frac{H_{l}}{E_l}. \label{eq:Cl}
\end{equation}
where the angular brackets mean averaging and $H_{l}$ is
the cross helicity on this scale. Thus, $H_l\approx C_l E_l$
and the substitution in Eq.~(\ref{eq:chi}) gives
\begin{equation}
C_l \delta v_l^3 \approx \chi l \xi_{l}^{(\chi)}. \label{eq:chi0}
\end{equation}
Comparing this expression with Eq.~(\ref{eq:en0}),
we can relate the correlation coefficient to the characteristics of the
exchange rate on the corresponding scale
\begin{equation}
C_l \approx \frac{\chi}{\varepsilon} \frac{\xi_{l}^{(\chi)}}{\xi_l}.
\label{eq:Cl1}
\end{equation}
If the delay in the helicity and energy exchanges
depends identically on the scale; i.e., $\xi_l^{(\chi)} \sim \xi_l$, the
velocity and magnetic-field correlation should be independent
of the scale (and, on the contrary, the correlation
independence of the scale means the identical
dependence of coefficients $\xi_l^{(\chi)}$ and $\xi_l$ on the scale).

In the strongly correlated turbulence, the relation
between the exchange times and the eddy turnover
time, which should be very large on the energy-transfer
scale, decreases with the scale approaching unity on the
dissipation scale. If it depends on the scale in the inertial
interval by the power law
\begin{equation}
 \xi_l  \approx  \xi_L \left({l / L} \right)^\mu,
\label{eq:qqq}
\end{equation}
the correction of the spectral energy distribution of pulsations
is unambiguously related to the parameter $\mu$
\begin{equation}
\delta v_l^2 \approx \varepsilon^{2/3} l^{2/3(1+\mu)}.
\label{eq:sp}\end{equation}

The conclusions about the role of cross-helicity in the stationary forced MHD turbulence are supported by results of numerical simulations using a shell model of MHD turbulence
Shell models describe the processes of spectral transfer
in the developed turbulence with the help of a limited
number of variables, each of which is a collective
characteristic of pulsations amplitudes of the velocity
field $U_n$ and the magnetic field $B_n$ in the wavenumber
interval $k_n<|\vec{k}|<k_{n+1}$, where $k_n=\lambda^n$ and $\lambda$ is the
interval (shell) width. The equations for collective variables are written to reproduce the "basic" properties of the initial equations of motion: the same kind of nonlinearity and integrals of motion.
The shell models are an
efficient tool for investigating the statistical properties
of developed small-scale turbulence (see, for example,
\cite{Frick2003}); in particular, they reproduce well the basic
known properties of MHD turbulence and the small scale
dynamo \cite{FS98}. However, model \cite{FS98} inherited the
basic disadvantage of cascade models associated with
the method of describing the helicity (in these models,
the different-sign helicity is attributed to shells
with even or odd numbers $n$). In this work, we used a
new model, which is obtained by a generalization on the
MHD case of the model proposed in \cite{Melander} for the helical hydrodynamic turbulence.
The model equations have the form
\begin{eqnarray}
 d_t U_n = i k_n (\Lambda_n(U,U)-\Lambda_n(B,B)) -
\frac{k_n^2 U_n}{\Ru} + f_n , \\ d_t B_n = i k_n
(\Lambda_n(U,B)-\Lambda_n(B,U)) - \frac{k_n^2 B_n}{\Rm},
\label{eq_sm}\end{eqnarray} where
\begin{eqnarray}
\Lambda_n(X,Y)= \lambda^2 (X_{n+1}Y_{n+1}+X_{n+1}^*Y_{n+1}^*)
-X_{n-1}^r Y_n \nonumber \\ -X_n Y_{n-1}^r+\imath\lambda(2
X_n^*Y_{n-1}^i+X_{n+1}^r Y_{n+1}^i-X_{n+1}^i Y_{n+1}^r) \nonumber
\\
+X_{n-1}Y_{n-1}+X_{n-1}^*Y_{n-1}^* -\lambda^2(X_{n+1}^r Y_n
\nonumber +X_n Y_{n+1}^r) \\+\imath\lambda(2
X_n^*Y_{n+1}^i+X_{n-1}^r Y_{n-1}^i-X_{n-1}^i Y_{n-1}^r), \nonumber
\end{eqnarray}
the asterisk designates conjugation, while the superscripts
$r,i$ are the real and imaginary parts. Without
dissipation, the total energy $E^T=\sum(|U_n|^2+|B_n|^2)/2$, the
cross helicity $H=\sum(U_nB_n^*+B_nU_n^*)/2$, and the magnetic
helicity $H_m=\sum k_n^{-1}((B_n^*)^2-B_n^2)/2$. If the magnetic
field is zero, the hydrodynamic helicity $H_h=\sum k_n((U_n^*)^2-U_n^2)/2$
is conserved. The distinctive feature
of the model is the possibility of the occurrence of
arbitrary-sign helicity in an arbitrary wavenumber
interval.
In all simulations, the Reynolds number and the
magnetic Reynolds number $\Ru=\Rm=10^6$, and the
shell width  $\lambda=1.618$. Time is measured
in dimensionless units equal to the eddy turnover
time on the maximal scale. The force $f_n$ operates
only in the two highest shells (the larger scales) providing
constant input of kinetic energy $\varepsilon=1$ and
cross helicity $\chi$.
In Fig.~1, we show the time-average values of the
total energy of the system obtained for various values of
$\chi/\varepsilon$ and agreeing well with estimate (\ref{eq:En(hi)}) to which the
solid line in the figure corresponds. In the same figure,
we show how the total energy $E^T$ of the system varies
with time for various levels of cross helicity input rate.
At $\chi=0$, the time of attaining the quasi-steady
state amounts to several vortex revolutions and, at $\chi=0.6$,
exceeds 100 dimensionless time units. In this case,
both the average value of energy and the character of its
oscillations vary.

\begin{figure}
\includegraphics[width=60mm]{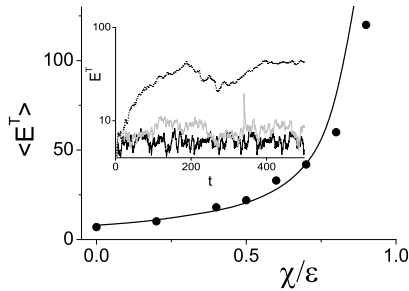}
\caption{Dependence of the average energy of the steadily
excited MHD turbulence on the injected cross helicity
$\chi/\varepsilon$. In the inset, the energy evolution is shown at $\chi=0$ (the
black line), $\chi=0.3$ (the gray line), and $\chi=0.6$ (the dotted
line).}
\end{figure}

The energy accumulation is also accompanied by
cross-helicity accumulation. In Fig.~2, we show the
average values of integral correlation coefficient $C=<H>/<E^T>$.
It is substantial that the turbulence accumulates
it at a low level of the injected cross helicity ($\chi/\varepsilon<<1$)
; i.e., the integral correlation coefficient
greatly exceeds the ratio between the injected helicity
and the injected energy. Thus, this tendency is
inherent not only to the Alfven turbulence \cite{gpl83}, but also
to the isotropic (Kolmogorov) MHD turbulence. At
large values of $\chi/\varepsilon$, the coefficient $C$ tends to unity.
Figure 3 shows how the energy spectra vary with
increasing the level of the cross helicity injected in
the flow. We present the energy values for each scale
compensated on the quantity $k_n^{2/3}$. In such a representation,
it is the horizontal line that corresponds to the
spectrum $k^{-5/3}$. It can be seen that there is such a spectral
range at $\chi=0$, while both the energy of each scale and
the spectrum slope increase with $\chi$.
It is of interest to trace directly the variation of the exchange time. In Fig.~4, we show the vortex turnover
time and the energy and helicity-exchange times calculated
for each shell in the turbulence with a high
level of cross helicity ($\chi/\varepsilon = 0.6$). It is indicative that the
exchange time on the integral scale exceeds the vortex turnover
time by almost two orders of magnitude.
This difference decreases with increasing the wave
number and vanishes in the dissipative range. In the
inertial range, the energy flux is also constant and the
exchange time is unambiguously determined by the
energy of pulsations of this scale; i.e., $t_n \sim <u_n^2>$. This
means that the power law for $t_n$ coincides with the slope
for the energy of pulsations. In the case shown in Fig.~4,
$t_n \sim l_n^{0.89\pm 0.02}$, and the energy distribution in the inertial
interval follows the law $<u_n^2> \sim l_n^{0.88\pm 0.02}$ (the straight
line in Fig.~3). The unexpected result is that the power
law for the exchange time is retained also in the dissipative
interval (Fig.~4). The helicity-exchange time
behaves similarly to the energy-exchange time, but it is
always somewhat less.
\begin{figure}
\includegraphics[width=60mm]{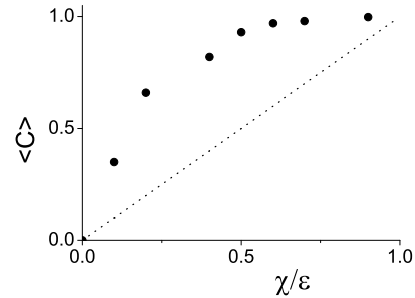}
\caption{Dependence of the average correlation level $C=<H>/<E>$
on the injected cross-helicity level $\chi/\varepsilon$.}
\end{figure}

\begin{figure}
\includegraphics[width=60mm]{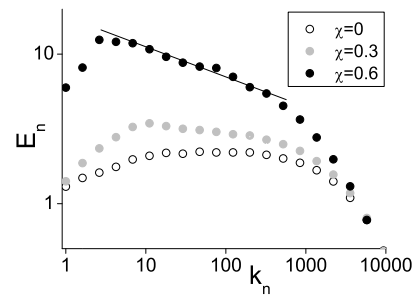}
\caption{Compensated energy spectra.}
\end{figure}

\begin{figure}
\includegraphics[width=60mm]{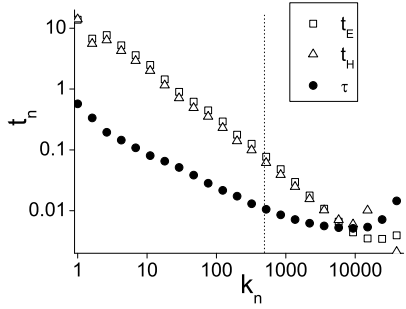}
\caption{Vortex-turnover and exchange times for the case of
$\chi/\varepsilon = 0.6$. The vertical line corresponds to the inertial-interval}
\end{figure}

Thus, it is shown that the cross helicity blocks the spectral
energy transfer in MHD turbulence and results in
energy accumulation in the system. This accumulation
proceeds until the vortex intensification compensates
the decreasing efficiency of nonlinear interactions.
The formula for estimating the average turbulence
energy is obtained for the set ratio between the injected
helicity and energy. It is remarkable that
the turbulence accumulates the injected cross helicity at its low rate injection -- the integral correlation coefficient
significantly exceeds the ratio between the injected
helicity and the energy. It is shown that the spectrum
slope gradually increases from "5/3" to "2" with the cross helicity
level.

\begin{acknowledgments}

This work was supported by the Russian Foundation
for Basic Research, no. 07-01-96007-Ural, and by a
grant of the Ural Division, Russian Academy of Sciences.
\end{acknowledgments}

\end{document}